\documentstyle[twoside,ijmpa1,epsfig]{article}
\def\qed{\hbox{${\vcenter{\vbox{			%HOLLOW SQUARE
   \hrule height 0.4pt\hbox{\vrule width 0.4pt height 6pt
   \kern5pt\vrule width 0.4pt}\hrule height 0.4pt}}}$}}
\begin{document}
\include{rlfig}
\runninghead{Associated top-Higgs production at future colliders}
{Associated top-Higgs production at future colliders}
\normalsize\textlineskip
\thispagestyle{empty}
\setcounter{page}{1}
\copyrightheading{}			%{Vol. 0, No. 0 (1993) 000--000}

\vspace*{0.88truein}

\fpage{1}
\centerline{\bf ASSOCIATED TOP-HIGGS PRODUCTION AT FUTURE COLLIDERS}
\vspace*{0.37truein}
\centerline{\footnotesize Sally Dawson}
\vspace*{0.015truein}
\centerline{\footnotesize\it Physics Department, Brookhaven National 
Laboratory}
\baselineskip=10pt
\centerline{\footnotesize\it Upton, NY 11973, USA}
\vspace*{10pt}
\centerline{\footnotesize Laura Reina}
\vspace*{0.015truein}
\centerline{\footnotesize\it Physics Department, Florida State University}
\baselineskip=10pt
\centerline{\footnotesize\it Tallahassee, FL 32306, USA}
%\vspace*{0.225truein}
%\publisher{(received date)}{(revised date)}

\vspace*{0.21truein}
\abstracts{We discuss the relevance of the associated production of a
Higgs boson with a pair of top-antitop quarks at both the LHC and a
future high energy $e^+e^-$ collider.}{}{}

\textlineskip			%) USE THIS MEASUREMENT WHEN THERE IS
\vspace*{12pt}			%) NO SECTION HEADING
%
%\vspace*{1pt}\textlineskip	%) USE THIS MEASUREMENT WHEN THERE IS
%\section{General Appearance}	%) A SECTION HEADING
%\vspace*{-0.5pt}
\noindent
\section{Overview and motivations}
The present and next generation of colliders will help elucidating the
nature of the electroweak symmetry breaking and the origin of fermion
masses. Lower bounds on the Higgs mass have been placed by LEP II:
$M_{h_{SM}}>113.2$ GeV for the Standard Model (SM) Higgs, and
$M_{h^0,A^0}>90.5$ GeV for the light scalar and pseudoscalar SUSY
Higgs\cite{osaka}. At the same time, precision fits of the Standard
Model seem to indirectly point at the existence of a light Higgs boson
($M_{h_{SM}}<170-210$ GeV)\cite{osaka}, while the Minimal
Supersymmetric Standard Model (MSSM) requires the existence of a
scalar Higgs lighter than about 130 GeV. Therefore, the possibility of
a Higgs discovery in the mass range around $120-130$ GeV seems around
the corner. If this does not happen by the Run II of the Tevatron, and
if the Higgs mechanism is the way the electroweak symmetry is broken,
almost certainly the LHC will discover it.
 
In this context the production of a Higgs boson (both SM and MSSM) in
association with a pair of top-antitop quarks is of extreme interest
for two reasons. First, the $t\bar tH$ production mode can be
important for discovery of a Higgs boson around $120-130$ GeV at the
LHC\cite{atlasreport}, and even at the Run II of the Tevatron with
high enough luminosity, as recently suggested\cite{davidetal}. Second, this production mode offers a direct handle
on the Yukawa coupling of the top quark, supposedly the most relevant
one to understand the nature of fermion masses. The Run II of the
Tevatron will not have enough statistics to use this feature, but both
the LHC and in particular a future high energy $e^+e^-$ collider will
be able to try a precision measurement of the $t\bar tH$ coupling.

In view of the role that this production mode can play in discovering
and studying the nature of a Higgs boson, we need to improve its
theoretical prediction and start developing more and more realistic
simulations. In particular it becomes crucial to estimate the
feasibility of a precision measurement of the $t\bar tH$ coupling at
the LHC and to compare it with the reach of a high energy $e^+e^-$
collider.

\section{Associated top-Higgs production at a high energy 
\boldmath$e^+e^-$\unboldmath collider}

\begin{figure}[hbtp]
\centering
\epsfysize=2.in
\leavevmode\epsffile{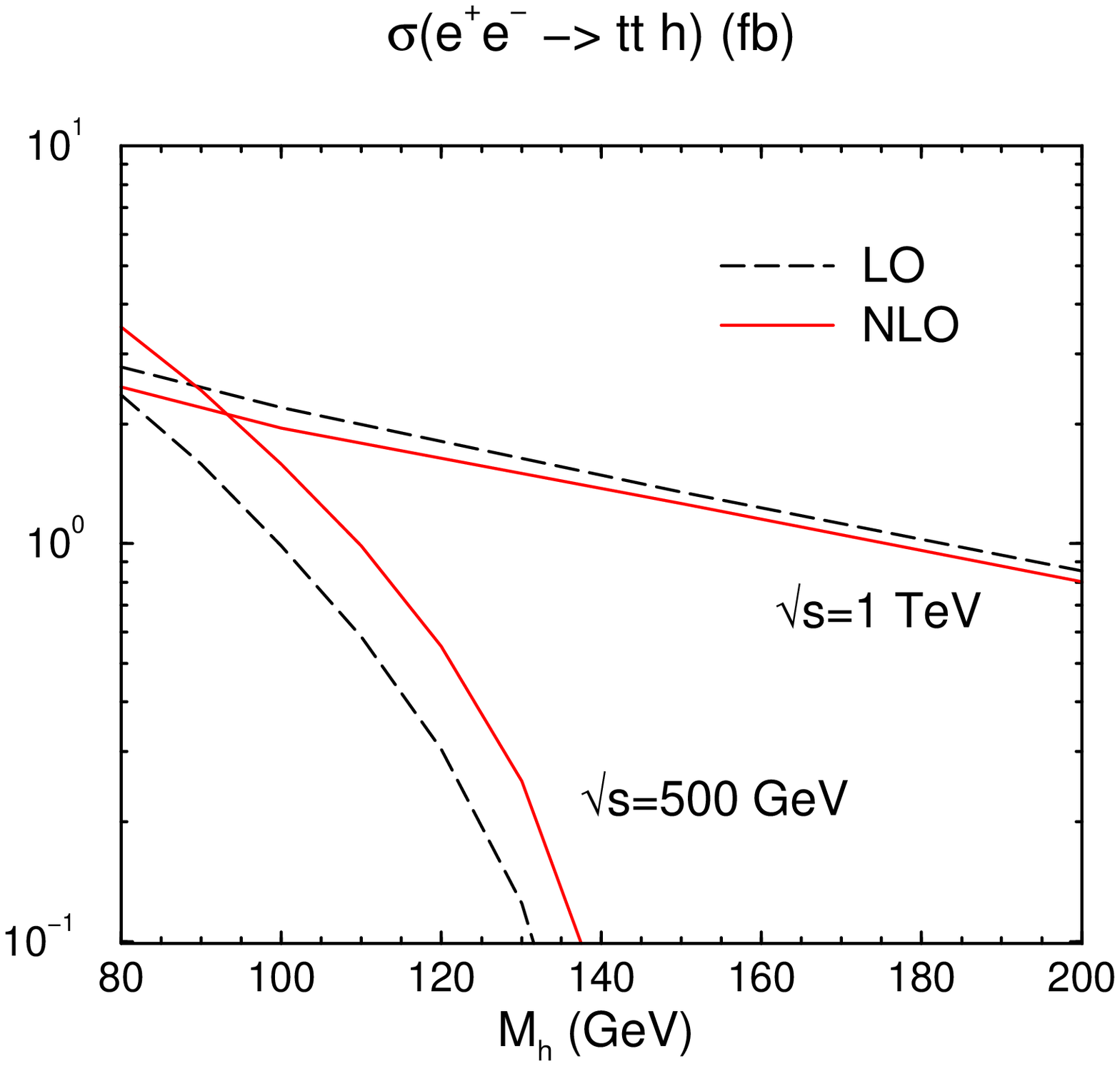}
\hspace{1.truecm}
\epsfysize=2.in
\leavevmode\epsffile{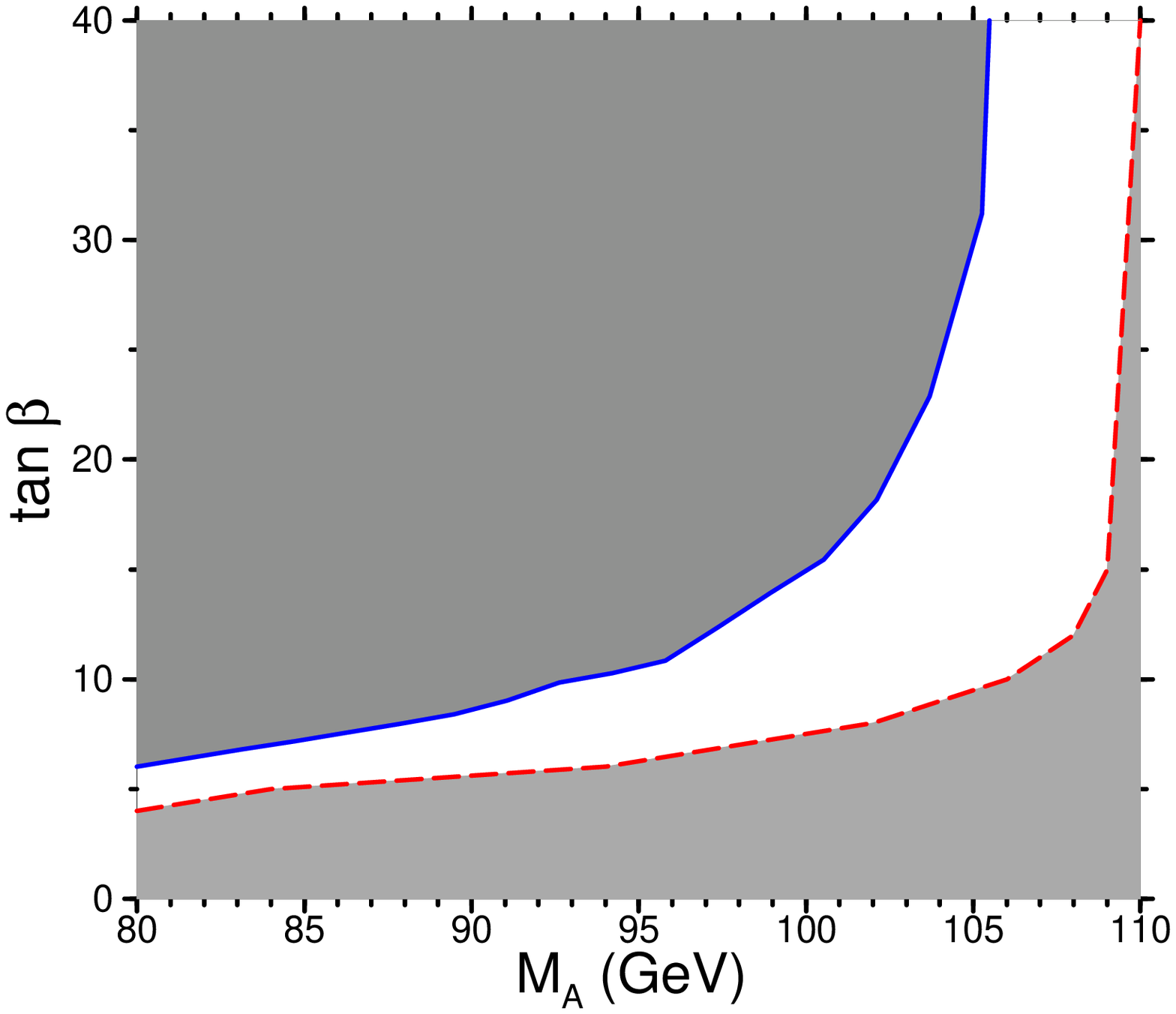}
\caption[]{\emph{Left plot}: the $O(\alpha_s)$
QCD cross section is compared to the lowest order cross section.  In
both cases, curves are shown for both $\sqrt{s}=500$~GeV and
$\sqrt{s}=1~\mbox{TeV}$. \emph{Right plot}: Regions in the
$M_A-\tan\beta$ plane where the cross section for $e^+e^-\rightarrow t
{\overline t} h_i^0$, ($h_i^0\!=\!h^0,H^0$), production is larger than
$0.75$~fb at $\sqrt{s}\!=\!500$~GeV.  The upper left hand region
results from $e^+e^-\rightarrow t {\overline t} H^0$, while the region
at the lower right is the result from $e^+e^-\rightarrow t {\overline
t} h^0$.  All NLO QCD corrections are included. The squarks are taken
to have a common mass, $M_S=500~GeV$ and and we assume no-mixing.  }
\label{fig:sm_susy}
\end{figure}

The process $e^+e^-\rightarrow t\bar t H$ has been studied quite
extensively in the last couple of years. The cross section turns out
to be highly sensitive to the top Yukawa coupling, both in the SM and
in the MSSM, over most of the parameter space. It has been calculated
both in the SM and in the MSSM at
$O(\alpha_s)$\cite{us,zerwasetal}. The main source of theoretical
uncertainty remains the scale dependence in $\alpha_s(\mu)$, which is
however below 10\%. For a SM Higgs, the factor
$K=\sigma_{NLO}/\sigma_{LO}$ at $\sqrt{s}\!=\!500$ GeV is in the range
$(1.4-2.4)$, depending on $M_H$. However the cross section is
drastically suppressed by phase space and for $M_H\!\simeq\!120-130$
GeV is of the order of $0.1\,fb$. On the other hand, at
$\sqrt{s}\!=\!1$ TeV the cross section for $M_H\!=\!120-130$ GeV is 
about $2\, fb$ and is only slightly reduced by QCD corrections
($K\!=\!0.8-0.9$).  Similar results holds in the MSSM case, where the
main channels are the scalar ones, i.e. $t\bar th^0$ and $t\bar tH^0$,
since the pseudoscalar mode $t\bar tA^0$ is very suppressed over most
of the MSSM parameter space. Both the SM and the MSSM results are
illustrated in Figure \ref{fig:sm_susy}.

Although the cross section is very small, the signature for $t\bar t
H$ production is spectacular. The possibility of fully reconstructing
the two top quarks in the final state allows to better discriminate
the signal over the background, and, together with a good b-tagging
efficiency, is crucial in increasing the precision with which the top
Yukawa couplings ($g_{ttH}$) can be measured at a high energy $e^+e^-$
collider. A first qualitative analysis\cite{us_analysis} for a
Standard Model Higgs boson has indicated that it will be very hard to
get a precise measurement at $\sqrt{s}\!=\!500$ GeV, even at high
luminosity, given the very limited statistics. However precisions of
the order of $7-15\%$ (statistical error only) are reachable at
$\sqrt{s}\!=\!1$ TeV for $M_H$ around 120-130 GeV (and $H\rightarrow
b\bar b$), assuming a b-tagging efficiency between 0.6 and 1.

It is interesting to observe that the optimal center of mass energy
for this process is neither $\sqrt{s}\!=\!500$ GeV nor
$\sqrt{s}\!=\!1$ TeV, but some scale in between. Here the impact of QCD
corrections is mild as for the $\sqrt{s}\!=\!1$ TeV case.  A detailed
simulation\cite{juste} of $e^+e^-\rightarrow t\bar tH$ for a SM Higgs,
at $\sqrt{s}\!=\!800$ GeV, has found that the top Yukawa coupling can
be measured with a precision of 5.5\%, when optimal b-tagging
efficiency is assumed.

\section{Associated top-Higgs production at the LHC}

\begin{figure}[hbtp]
\centering
\epsfysize=1.8in
\leavevmode\epsffile{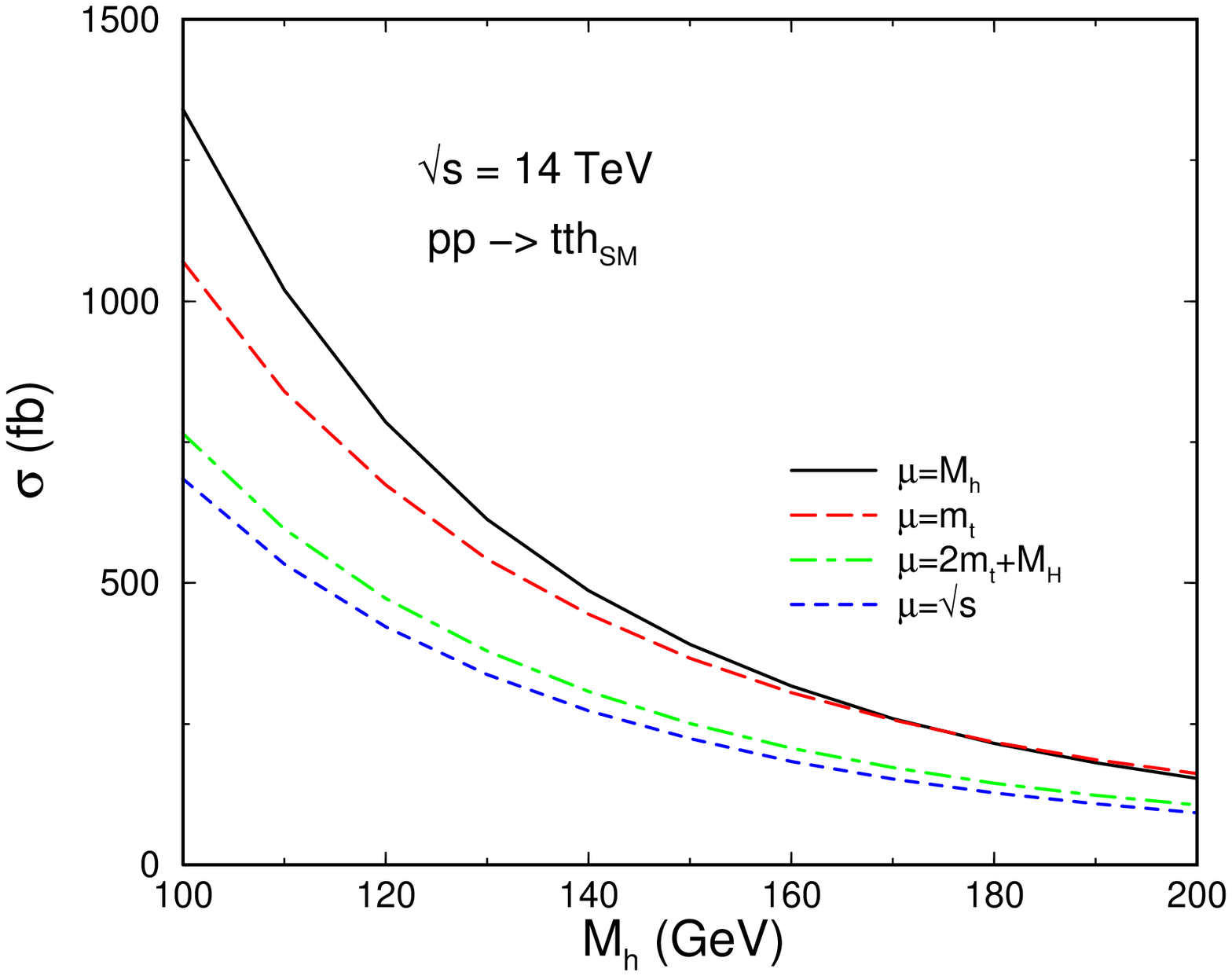}
\hspace{1.truecm}
\epsfysize=1.8in
\leavevmode\epsffile{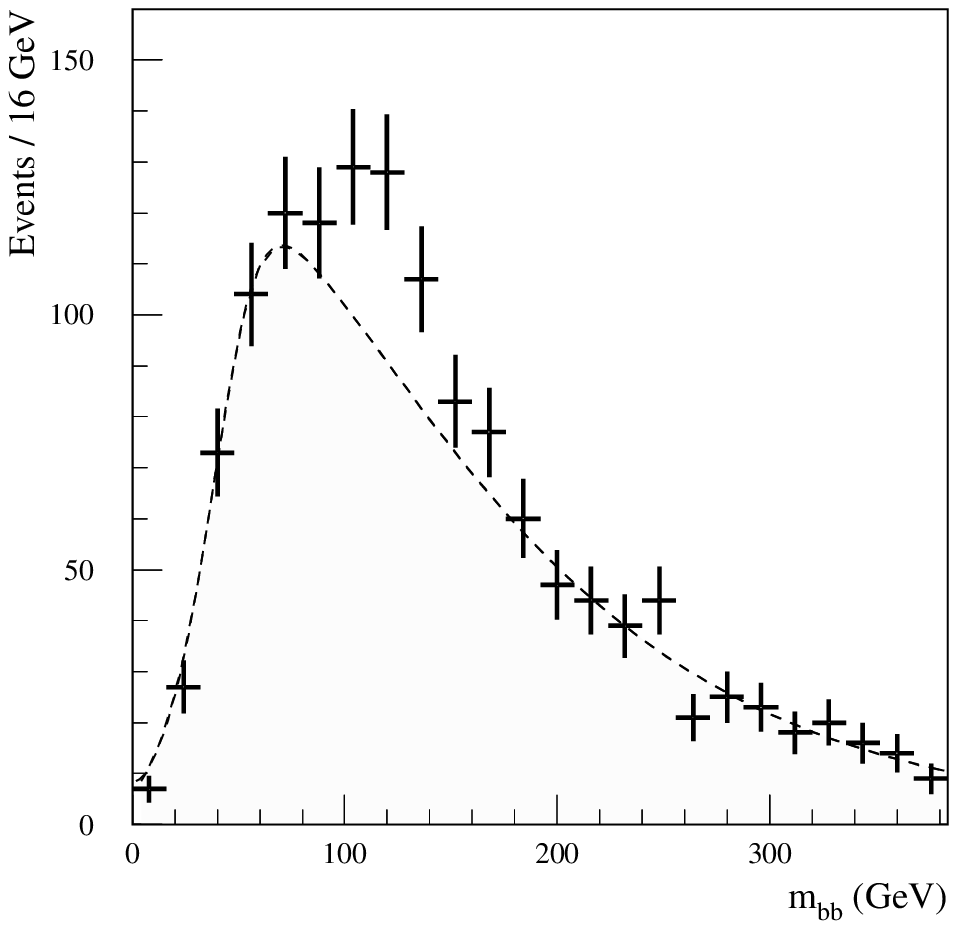}
\caption[]{\emph{Left plot}: Cross
section for $t\bar th_{SM}$ production at the LHC ($\sqrt{s}\!=\!14$~TeV)
as a function of the Higgs mass, for several values of the scale $\mu$.
\emph{Right plot}: Invariant mass distribution of tagged $b$-jet pairs in fully
reconstructed $t\bar th_{SM}$ signal events and background events,
obtained using the fast simulation of the ATLAS detector, for
$m_{h_{SM}}\!=\!120$~GeV and integrated luminosity of $100~fb^{-1}$
($30~fb^{-1}$ for low luminosity and $70~fb^{-1}$ for high). The
points with error bars represent the result of a single experiment and
the dashed line represents the background distribution.}
\label{fig:lhc}
\end{figure}

The cross section for $p\bar p\rightarrow t\bar tH$ or $pp\rightarrow
t\bar tH$ is currently known only at the tree level\cite{kunszt}, and
is therefore affected by a very strong scale dependence, as can be
seen in Figure \ref{fig:lhc} for the case of a SM Higgs. The
calculation of the $O(\alpha_s)$ QCD correction is work in progress by
several groups.

The existing analyses have been performed for a Standard Model like
Higgs and have assumed a small theoretical uncertainty, as we expect
to be the case by the time both the Run II of the Tevatron and the LHC
turn on.  The relevance of the $t\bar tH$ channel for Higgs discovery
at the Tevatron has been discussed in a parallel session of this
meeting\cite{davidetal}. For the LHC, most of the analyses have been
performed by the ATLAS collaboration\cite{atlas,sapinski} and have
been recently summarized in the context of an LHC workshop\cite{lhc},
to which we refer for full details.  It has been shown that, within
the Standard Model, $t\bar tH\,(H\rightarrow b\bar b)$ is among the
most important channels for discovery of a low mass mass Higgs
($M_H\simeq 100-130$ GeV). In this channel it is possible to obtain a
quite large signal significance and also to measure the top Yukawa
coupling.  An example of the signal that can be obtained in the
invariant $b\bar b$ mass distribution for $100\,fb^{-1}$ of integrated
luminosity is shown in Figure \ref{fig:lhc}. For the same set of
parameters and integrated luminosity, and assuming a separate
determination of the $b\bar b H$ Yukawa coupling, the top Yukawa
coupling can be determined with a precision of about 16\%.  Further
analyses which include more Higgs decay channels will very likely
improve this precision.

\end{document}